\documentclass[aps,prd,groupedaddress,nofootinbib,letterpaper]{revtex4}

\usepackage{amsmath}
\usepackage{amsfonts}
\usepackage{amssymb}
\usepackage{amsthm}
\usepackage{graphicx}
\usepackage[utf8]{inputenc}

\usepackage{braket,mathtools}
\usepackage{mathrsfs}
\usepackage{stmaryrd}

\usepackage{hyperref}
\usepackage{color}

\newcommand{\cala}{{\cal A}}
\newcommand{\call}{{\cal L}}
\newcommand{\calo}{{\cal O}}

\newcommand{\beq}{\begin{equation}}
\newcommand{\eeq}{\end{equation}}
\newcommand{\bea}{\begin{eqnarray}}
\newcommand{\eea}{\end{eqnarray}}

\begin{document}
\begin{titlepage}

\title{Exploring strong-field deviations\\ from general relativity via gravitational waves}

\author{Steven B. Giddings}
\email{giddings@ucsb.edu}
\affiliation{Department of Physics, University of California, Santa Barbara, CA 93106}
\affiliation{CERN, Theory Department,
1 Esplande des Particules, Geneva 23, CH-1211, Switzerland}

\author{Seth Koren}
\email{koren@physics.ucsb.edu}
\author{Gabriel  Trevi\~no}
\email{gabriel@physics.ucsb.edu}
\affiliation{Department of Physics, University of California, Santa Barbara, CA 93106}

\begin{abstract}
Two new observational windows have been opened to strong gravitational physics:  gravitational waves, and very long baseline interferometry.  This suggests observational searches for new phenomena in this regime, and in particular for those necessary to make black hole evolution consistent with quantum mechanics.  We describe possible features of ``compact quantum objects" that replace classical black holes in a consistent quantum theory, and approaches to observational tests for these using gravitational waves.  This is an example  of a more general problem of finding consistent descriptions of deviations from general relativity, which can be tested via gravitational wave detection.  Simple models for compact modifications to classical black holes are described via an effective stress tensor, possibly with an effective equation of state.  A general discussion is given of possible observational signatures, and of their dependence on properties of the colliding objects.  The possibility that departures from classical behavior are restricted to the near-horizon regime raises the question of whether these will be obscured in gravitational wave signals, due to their mutual interaction in a binary coalescence being deep in the mutual gravitational well.  Numerical simulation with such simple models will be useful to clarify the sensitivity of gravitational wave observation to such highly compact departures from classical black holes.

\end{abstract}

\maketitle

\end{titlepage}

\section{Introduction}

With a steadily increasing number of gravitational wave observations from coalescing binaries\cite{LIGOScientific:2018mvr}, and with imminent new data from very long baseline interferometric (VLBI) observations of apparent black holes\cite{Doeleman:2009te}, we have entered a new era of observationally testing strong-field gravity.  As it endures increasingly precise tests in this strong field realm, general relativity (GR) so far appears to be holding firm.  

Modifications to the classical behavior of black holes are of course possible, and many models with such modified behavior have been considered, with various motivations.  But there is one very compelling reason to believe that a classical description of black holes {\it must} ultimately be modified:  such a description appears inconsistent with quantum mechanics, which is thought to govern all physical phenomena.  At first it was believed that this might only be important very near the center of a black hole, or in the late stages of black hole evaporation, and would be irrelevant outside the horizon of large black holes.  However, deeper examination of the requirement of consistency of black hole evolution with quantum mechanics has led to a widespread view\cite{BHMR,LQGST,BHIUN,Mathurrev,NLvC,AMPS,MaSu} that there need to be important corrections to classical black hole behavior {\it at horizon scales}.  This is due to the requirement that for ultimate unitarity of black hole evolution, information needs to transfer out of a black hole while it is still of macroscopic size, in direct contradiction with a description based on classical geometry, together with small perturbations due to local quantum fields.  

This raises a key question:  can the modifications to classical behavior necessary for quantum consistency have observable effects?\footnote{For some earlier discussion of this question, see \cite{SGobswind,SGGW,SG-nature}.  For other discussion of tests for deviations from GR for black holes, see {\it e.g.}  \cite{Yagi:2016jml,Cardoso:2017njb,Barack:2018yly,Carballo-Rubio:2018jzw}.} Of course, often when physics has opened new observational regimes, new phenomena have been found, so it is important to model and investigate possible new phenomena in the strong gravity regime, independent of this question.  But, quantum consistency serves as a particularly important motivator to focus on the specific class of effects that can restore consistency between the existence of black hole-like objects and quantum mechanics.

This leads to a particular focus on the near-horizon region, $r\sim 2M$ for a black hole without spin.  The origin of the unitarity problem (in some views, really a crisis) is in the physics of Hawking radiation, which can be thought of as originating near\footnote{For recent discussion, see \cite{Giddings:2015uzr}.} $r\sim 3M$.  Quantum consistency strongly motivates the possibility of new quantum effects in this region, or closer to the horizon\cite{AMPS}.  

A variety of scenarios have been considered for near-horizon modifications necessary for quantum consistency.  The resulting objects, whose description is supposed to be consistent with quantum mechanics, have varying degree of departure from the classical black hole (CBH) description, depending on the scenario.  In order not to prejudice a particular scenario at the outset, but given their quantum origin, we will refer to these objects generically as {\it compact quantum objects} (CQOs).\footnote{A related terminology is exotic compact object (ECO).  A CQO is intended to be something more specific than an ECO,  since CQOs are presumed to owe their existence to the quantum dynamics necessary to make quantum evolution consistent for CBH-like objects -- and quantum mechanics is certainly not exotic.}

A central question then becomes what non-CBH properties CQOs have, and how these might be detected.  While VLBI is expected to provide an important window,\footnote{Some discussion of this appears in \cite{SGobswind,GiPs,SG-nature}.} this paper will focus on gravitational waves.  In order to predict the gravitational wave signature of coalescence resulting from a particular scenario or model, one needs to describe the full nonlinear evolution of the binary, analogous to the nonlinear evolution of GR.  This {\it problem of nonlinear evolution} is a first challenge.  This is particularly true given that while existing models for CQOs exhibit some of the behavior important for quantum consistency, they are not yet derived from a more complete underlying theory of quantum gravity.  
This is an example of a more general problem in observationally testing GR -- the lack of good foils or alternatives to GR for describing nonlinear evolution of alternatives to CBHs, in the strong-gravity regime.

A second challenge for tests of  highly compact objects is what we call the {\it problem of gravitational obscuration}.  Suppose we consider the collision of two objects that differ from CBHs only in a region close to the horizon.  We might in general expect there to be little deviation from GR in the gravitational wave signal until the  discrepant regions meet.  However, this happens when the two objects are deep inside a gravitational well and we might expect that if the objects coalesce to form an object with horizon-like behavior, most of the discrepant signal is also absorbed into this final  object and is not observed at infinity\cite{SGGW}.  This potentially considerably lowers observational sensitivity.

Of course, it may be that there are signatures of CQO properties from inspiral, and CQO behavior may also affect observations by changing absorption and reflection probabilities for gravitational waves, but these may be more subtle effects.

Note that this discussion contrasts with another proposal for modifications to the CBH signal, that of gravitational echoes\cite{Cardoso:2016rao,Cardoso:2016oxy,Abedi:2016hgu}.  The key difference  arises from the fact that the echo story assumes that the two objects coalesce to immediately form an object that does not have horizon-like behavior, {\it e.g.} by having or rapidly developing a ``hard" barrier, from which the echoes reflect.  This represents a more extreme departure from CBH behavior than appears to be required for quantum consistency.  We instead focus on the possibility that the effects needed for a reconciliation with quantum mechanics involve less drastic departure from CBH behavior.

A goal of this paper is to begin work to investigate these related problems, in some simple models for departure from CBH behavior.  In fact, a first model for a departure from a CBH merger is a merger of neutron stars, whose gravitational signatures exhibit important features.\footnote{See, {\it e.g.}, \cite{Palenzuela:2015dqa}.} Of course neutron stars cannot have the masses seen in many recent detections, so more general models are needed.  But this suggests one general approach to investigating departures from CBHs that have consistent nonlinear evolution, namely to parametrize them in terms of an effective stress tensor, and in even simpler models, in terms of an effective equation of state (EOS).  Such models, if they produce objects with relevant masses, begin to provide simple foils for coalescence of CBHs. 

After  further discussion of motivation and CQO scenarios in the next section, section III will describe such an effective approach and its use to formulate simple models to test aspects of possible modifications to CBH behavior, such as with CQOs.   Section IV will investigate spherically-symmetric solutions for such models, and in particular those with an EOS that permits them to be highly compact, in line with preceding comments.  Section V will discuss parameters and features of such solutions, and their possible connection to observable deviations in gravitational wave signatures.  While the models we study are limited in their ability to capture possible CQO properties, they should allow initial investigation of some of the basic questions regarding gravitational wave sensitivity to very compact departures from CBHs.

In particular, one ultimate goal is to understand how sensitive gravitational wave observations can be to highly compact deviations from CBHs, given the obscuration question. The models we describe provide a way to set up the problem, but probably the best way to test the role of obscuration -- and other aspects of sensitivity to highly-compact deviations -- is through numerical evolution of the kinds of solutions that we provide.  This is an important step for future work.

In short, now that strong gravitational physics is an observational subject, it is important to try to parameterize possible deviations from the predictions of GR, and to test them against observation.  This paper will begin to investigate some models for certain kinds of deviation from classical black hole behavior, in an effective approach.

\section{CQO models and observational challenges}\label{motiv}

Gravitational dynamics has been quantitatively well tested primarily in weak-field regimes, perturbatively close to flat space (for a recent review, see \cite{Will:2018lln}).  This leaves as an important question the possibility of deviations from GR in strong-field regimes.  GR is a mathematically beautiful and compelling framework, and  it is commonly believed that it will be significantly modified only in strong {\it curvature} regimes.  However, there are forceful arguments for some modification of the combined frameworks of GR and local quantum field theory (LQFT) in situations where classical GR predicts a black hole horizon would form.  The vicinity of such horizons can also be thought of as strong-field regions;  although for big black holes they are not expected to have high curvatures, the metric near a black hole horizon corresponds to a large perturbation  of  an ambient Minkowski space in which the black hole resides.

A primary motivation to expect such modification is the combination of facts: 1) black hole-like objects appear to exist and 2) attempts to describe black holes as objects in a quantum Universe, based on a combination of LQFT and GR, appear to produce a contradiction with  basic principles of quantum mechanics.  After much exploration of this ``information paradox" or ``unitarity crisis," many who have thought deeply about this puzzle have concluded that modifications to GR+LQFT are required, not just at very short distances, but, in the context of black holes, at scales given by the horizon size, which can be arbitrarily large for big black holes.  This appears to be the most conservative approach to reconciling the existence of compact objects that have basic features of classical black holes with quantum mechanics.

A number of proposals have been considered for modifications to the GR+LQFT description at a scale given by the radius $R$ of a classical black hole (BH) horizon, or at even larger scales, while respecting quantum principles.  These can be divided into some broad scenarios for CQOs:

\begin{enumerate}

\item {\it Massive remnants.}  A very general scenario was proposed in \cite{BHMR}: at some stage in its evolution, a BH  transitions to a new kind of star-like ``massive remnant," truncating the Schwarzschild spacetime outside the would-be horizon, analogous to, {\it e.g.},  a neutron star.\footnote{A variant proposal is that such a massive remnant forms before a BH can form.}  In such a scenario the new physics outside the horizon is assumed to be characterized by some short (microscopic) distance scales, and thus be ``hard;"  one measure of this is typical momentum transfer to infalling matter.\footnote{Either the gravitational field may be characterized by hard scales, or other fields or structure may be characterized by such hard scales.}  A number of more specific variants of this basic hard picture have been proposed.  These include gravastars\cite{gravast}, fuzzballs\cite{Mathurrev}, firewalls\cite{AMPS}, and Planck stars\cite{RoVi}.  These may differ in their evolution subsequent to the transition.  

\item {\it Soft gravitational atmospheres.}  Another possibility is  a ``softer" departure from CBHs, in a near-horizon ``atmosphere" region, which is similar to the near-horizon region of a CBH and in particular permits infalling observers to pass without undue violence\cite{NLvC,SGmodels,BHQIUE,NVNLT,NVU}.  The characteristic softness scale should be determined by  a distance scale that at the least grows with black hole radius.  The deviations present in the atmosphere are constrained by the fact that interactions with them must suffice to transfer information from the BH state to the environment of the BH, so that evolution is unitary.  If these interactions are assumed to couple universally to all fields, two variants have been described.  One is ``strong" \cite{NVNLT}, with an effective description of the atmosphere in terms of $\calo(1)$ but soft state-dependent metric fluctuations.  A more minimal scenario is the ``weak" scenario of \cite{NVU}, in which very small state dependent metric fluctuations are found to be sufficient to transfer information.

\item {\it Long distance modification of locality.}  A third possibility is modification to the locality structure of LQFT on scales $\gg R$.  A standard example of this is the ER=EPR proposal\cite{vanR,MaSu}, where mere entanglement of remote degrees of freedom is interpreted as corresponding to formation of a connection between them via a spacetime bridge.

\end{enumerate}

In each of these scenarios for CQO behavior, a very interesting and important question is whether the required new effects could have any observational implications.  This question becomes even more compelling with growing prospects for testing strong field gravity, both through gravitational wave observation with LIGO/VIRGO, and in the future with LISA, and with VLBI, specifically with EHT.  

These two observational approaches have key differences.  VLBI effectively provides an electromagnetic picture of the geometry, resulting from passage of light from, {\it e.g.} accreting matter, through the region near the (would-be) horizon.  Thus, all that is required of a scenario is a prediction of how light propagates or interacts with the CQO replacement of a CBH.  For example, in the ``soft, strong" proposal of \cite{NVNLT}, one can perform ray-tracing through the perturbed geometry to determine possible modifications of images that could be visible to EHT\cite{GiPs}.  Similar predictions of electromagnetic images are in principle possible from any other sufficiently explicit scenario.  

As noted, tests of scenarios via gravitational waves (GWs) face the problem of nonlinear evolution; they require dynamics as opposed to simply providing ``snapshots" of the configuration. Specifically, the prediction of the GW signal requires prediction of the full nonlinear evolution of the CQOs, analogous to the nonlinear evolution of GR.  Although one or more of the broad scenarios outlined above may produce a logically-viable description of the quantum behavior, none of them is yet advanced enough to be based on an understanding of underlying gravitational dynamics that is sufficiently developed to make predictions about the nonlinear evolution of CQO replacements for CBHs.  

The second problem described above, obscuration,  is also a potentially important challenge to  using GWs as probes of novel structure that could modify a CBH.  Specifically, suppose a CBH is replaced by a CQO departing from CBH behavior out to a radius\footnote{For spherically symmetric objects, this can be precisely defined in terms of the standard Schwarzschild coordinate $r$, using the match to the Schwarzschild solution at $r>R_a$.} $R_a=R+\Delta R_a$. When two such objects collide after the end of inspiral,  one na\"\i vely expects significant modification to the GW signature from the regime where the modified structures come close to touching\cite{SGGW}.  But, if $\Delta R_a \lesssim R$, this occurs when the CQOs are deep within their mutual gravitational well, suggesting that a significant part of the signal modification may be absorbed in the final object, if it indeed has basic features of a CBH.  This seems particularly clear if $\Delta R_a\ll R$.  Indeed, there are arguments that much of the final GW signal from coalescence of BHs is generated in the vicinity of the light ring of the final BH\cite{Cardoso:2016rao} (though for some counterpoints, see \cite{Khanna:2016yow}).  So, an important question is to what extent GW observations can be sensitive to possible near-horizon quantum structure.  

Given the growing and anticipated amount of GW data and the lack of predictions of nonlinear evolution for complete quantum scenarios, one reasonable approach is to begin by exploring the questions of sensitivity of observations to new structure and obscuration in simple models for modification of CBH behavior.  Specifically, if a CBH is replaced by a CQO with different properties,  how much effect can this have on the GW signal -- how sensitive are GW observations to {\it any} modification of structure in the near-horizon region?  Since answering this question requires  nonlinear evolution of the CQOs, one simple way to begin to explore this question is to assume that whatever the full description is of their configuration, it can be approximately described as possessing an effective four-dimensional metric $g_{\mu\nu}$, and that departures from the vacuum Einstein equations can be consistently and approximately summarized by an effective stress tensor source, $T_{\mu\nu}$, in these equations.\footnote{In a full quantum theory, these are significant assumptions.}  If this stress tensor satisfies certain consistency conditions, such as conservation, this provides a model for how to incorporate modifications to CBH behavior with consistent nonlinear evolution.

Various specific models for departure from BH predictions for GW signatures have also been considered, including boson and fermion stars (see {\it e.g.} \cite{Giudice:2016zpa,Palenzuela:2017kcg,Bezares:2018qwa}) and Proca stars\cite{Sanchis-Gual:2018oui}.  However, the resulting solutions have characteristic sizes determined by mass parameters of the underlying theory, so do not provide  models for  quantum behavior of BHs of arbitrary size, and also do not typically achieve the highest range of compactness.   For this reason, we explore other forms of the stress tensor.

\section{Effective approach}

The complete quantum description of CQOs replacing BHs is so far unknown, and may at the fundamental level involve different quantum variables than a four-dimensional metric.  However, in order to test sensitivity to departures from classical GR, we will assume the existence of a quantum variable $g_{\mu\nu}$ that plays the role of an effective four-dimensional metric, such that $\langle g_{\mu\nu}(x) \rangle$ is well behaved (non-planckian) in a typical state.  This may or may not be true in various scenarios -- for example in the cases of fuzzballs or firewalls, if a consistent fundamental description even exists for either of these proposals in situations corresponding to large, non-extremal BHs.  
In either a general such massive remnant  scenario, or in that of a soft gravitational atmosphere, there may also be other quantum degrees of freedom that are excited in the vicinity of the would-be horizon.  We will model the effect of these, and of possible corrections to classical Einsteinian evolution, using an effective stress tensor,
\beq\label{Qevolve}
\langle G_{\mu\nu}[g] \rangle = 8\pi G \langle T_{\mu\nu}\rangle\ .
\eeq
If the underlying fundamental theory {\it were} a field theory, this would  arise from an action
\beq
S=\int d^4x \sqrt{-g} \left( \frac{R}{16\pi G} +\call\right)
\eeq
where $\call$ is a lagrangian summarizing other degrees of freedom as well as, for example, higher-curvature terms.  The effective stress tensor is then
\beq
T_{\mu\nu} = -\frac{2}{\sqrt{-g}} \frac{\delta}{\delta g^{\mu\nu}} \int d^4x \sqrt{-g} \call\ .
\eeq
While the origin of the quantum evolution law in a more fundamental description of quantum gravity may not be from such an action, {\it e.g.} in the case of a soft gravitational atmosphere, working with  an effective stress tensor gives an approach to test sensitivity to some types of near-horizon deviations from CBH behavior.  
Such an effective stress tensor must obey certain consistency conditions; one is conservation, 
\beq\label{EMcons}
\langle \nabla^\mu T_{\mu\nu}\rangle =0\ .
\eeq

We ultimately wish to study coalescence of two CQOs in such a model, governed by the evolution law \eqref{Qevolve}.  We begin by considering the description of the individual objects.  In the spherically-symmetric static case, with zero angular momentum, the stress tensor in spherical coordinates $x^\mu=(t,r,\theta,\phi)$ must take the form
\beq\label{ssstress}
\langle T^{\mu}_{\nu}\rangle = {\rm diag}[\rho(r), p_r(r), p_\theta(r), p_\theta(r)]\ .
\eeq
The underlying quantum dynamics then determine $\rho, p_r, p_\theta$, subject to conservation \eqref{EMcons}, as well as other consistency conditions.  To proceed further, we need more information about relations between these variables, and the metric.

If the properties of the CQO can be described in this fashion, at least approximately, the relations between $\rho, p_r, p_\theta$, and $g_{\mu\nu}$  would depend on the currently unknown full quantum dynamics. A goal of this paper is  to begin to investigate sensitivity to such dynamics, replacing CBH behavior in the near-horizon region.  Since a primary question is how much of the signal from colliding compact objects is absorbed into the final BH, we can start to explore this sensitivity by considering simplified  models for CQOs.   One highly simplified but pragmatic approach is to assume that the quantum stress tensor is well-approximated by \eqref{ssstress}, and moreover behaves like an isotropic ideal fluid with an equation of state,
\beq\label{EOS}
p_r=p_\theta=p(\rho)\ .
\eeq
In this case, one additional plausible consistency condition is that the speed of sound not exceed the speed of light, requiring
\beq\label{caus}
p'(\rho)\leq 1\ .
\eeq

For consistency with quantum mechanics, CBHs of arbitrary size must ultimately be replaced by CQOs.   As noted above, this is not true for various specific microscopic models for $\call$, such as  
boson stars (see \cite{Liebling:2012fv}, and references therein), which have maximum masses that depend on the mass of the boson that is coupled to GR to find non-trivial solutions.  A way to ensure the existence of solutions at arbitrary mass is if the effective dynamics has a scaling symmetry.

The basic scaling transformation replaces the metric configuration $g_{\mu\nu}(x)$ with a new configuration,
\beq\label{scalexm}
g_{\mu\nu}(x) \rightarrow {\tilde g}_{\mu\nu}(x) = g_{\mu\nu}(\lambda x)\ .
\eeq
This leaves the Minkowski metric invariant, but rescales perturbations of it:
\beq\label{pertrescale}
ds^2=[\eta_{\mu\nu} + h_{\mu\nu}(x)]dx^\mu dx^\nu \rightarrow d\tilde s^2=[\eta_{\mu\nu} + h_{\mu\nu}(\lambda x)]dx^\mu dx^\nu\ .
\eeq
For example, this transformation maps the Schwarzschild solution with mass $M$ to a solution with mass $M/\lambda$.
A change of integration variables shows that
\beq
{ S}[\tilde g]= \int d^4x \sqrt{-\tilde g} \tilde R =  \frac{1}{\lambda^2} \int d^4x \sqrt{- g} R= \frac{1}{\lambda^2}S[g]\ ,
\eeq
but of course the vacuum equations are invariant:
\beq
{\tilde G}_{\mu\nu}(x) = \lambda^2 G_{\mu\nu}{}_{\vert_{\lambda x}} =0\ .
\eeq

With a source, the equations will be scale invariant if the scaling transformation also acts as
\beq\label{Tscale}
T_{\mu\nu}\rightarrow {\tilde T}_{\mu\nu}(x)= \lambda^2 T_{\mu\nu}(\lambda x)\ .
\eeq
For example, the stress tensor of a massless scalar,
\beq
T_{\mu\nu}= \partial_\mu\phi\partial_\nu\phi -\frac{1}{2} g_{\mu\nu} (\partial \phi)^2
\eeq
satisfies this condition, but it is violated if there is a scalar mass present.

For the stress tensor \eqref{ssstress}, there will be static solutions of all scales if given a solution  $\rho(x), p_r(x), p_\theta(x)$, there is a solution 
$\lambda^2 \rho(\lambda x), \lambda^2 p_r(\lambda x), \lambda^2 p_\theta(\lambda x)$.  This is clearly violated by a given fixed equation of state, \eqref{EOS}.  Thus, to have solutions of all scales, one must consider a {\it family} of equations of state; this is more plausible if $\langle T_{\mu\nu}\rangle$ summarizes some general properties of CQOs, as opposed to being determined directly by a specific microphysical $\call$.  We return to this point later.

\section{Some simplified models for compact quantum objects}\label{Models}

New effects associated to a CQO replacement of a classical BH are 
plausibly only significant in the strong gravity region, near the would-be horizon of the classical BH.  As was noted above, this raises the question of gravitational obscuration of any modification to the classical GR signal resulting from the collision of two such objects.  In order to explore this question, we first explore possible compact solutions, in the effective approach outlined above.

\subsection{Generalities}

The general static, spherically-symmetric metric can be written
\beq
ds^2= - e^{2\mu(r)} dt^2 + e^{2\lambda(r)} dr^2 + r^2(d\theta^2 + \sin^2 \theta d\phi^2)\ ;
\eeq
we assume this form for the effective metric $\langle g_{\mu\nu}\rangle$.
One commonly introduces an effective mass $m(r)$ by 
\beq
e^{-2\lambda(r)}= 1-\frac{2m(r)}{r}\ .
\eeq
With the stress tensor \eqref{ssstress}, Einstein's equations \eqref{Qevolve} take the form (see, {\it e.g.}, \cite{Andreasson:2007ck})
\bea
m'&=& 4\pi r^2 \rho\\
 \mu'&=&\frac{1}{r}\left(4\pi r^2 p_r +\frac{m}{r}\right)e^{2\lambda}\\
\mu^{\prime\prime} + (\mu'-\lambda')\left(\mu'+\frac{1}{r}\right) &=& 8\pi p_\theta e^{2\lambda}\ ,
\eea
where we use units with $G=1$ and prime denotes the $r$ derivative.
The Tolman-Oppenheimer-Volkov (TOV) equation generalizes to 
\beq\label{TOVgen}
p_r'=-\mu'(p_r+\rho) -\frac{2}{r} (p_r-p_\theta) = -\frac{1}{r}(p_r+\rho) \left(4\pi r^2 p_r +\frac{m}{r}\right)e^{2\lambda} -  \frac{2}{r} (p_r-p_\theta)\ .
\eeq

If a solution of these equations has vanishing $\langle T^{\mu}_{\nu}\rangle$ outside a radius $R_a$, and has total mass $M$, we can define the compactness of the solution to be $C=M/R_a$.   Our focus is on sensitivity to merger of highly-compact solutions.  While it is possible to find more compact anisotropic solutions\cite{Andreasson:2007ck,Karmakar:2007fn} (see also \cite{Raposo:2018rjn,Kaplan:2018dqx} for recent discussions), we defer exploring these to future work and instead focus on the simpler isotropic case, $p_r=p_\theta=p$.   In that case, the preceding equations can be shown to imply that, for positive $p$ and $\rho$,  the compactness is limited by the Buchdahl bound\cite{Buchdahl:1959zz}, 
\beq\label{buchbd}
C<4/9\ .
\eeq

However, the Buchdahl bound is not  achievable using a physical equation of state (EOS).  Specifically, the stiffest EOS satisfying the constraint \eqref{caus} on the speed of sound is
\beq\label{mceos}
p=
\begin{cases}
\rho-\rho_0 , & \text{if}\ \rho>\rho_0 \\
      0, & \text{if}\ \rho<\rho_0\ ,
    \end{cases} 
\eeq
where $\rho_0$ is an EOS parameter.
It has been shown\cite{Fujisawa:2015nda} that this EOS yields the isotropic solutions with the highest compactness\cite{Lattimer:2012nd} satisfying the causality condition\eqref{caus}.\footnote{This EOS has been frequently studied for the purpose of determining a theoretical maximum mass and radius for neutron stars; see  {\it e.g.}  \cite{HaenselZdunik, BrecherCaporaso, RhoadesRuffini, KalogeraBaym, Glendenning, Korandaetal}.}  The surface of these solutions is the radius $R_a$ where $\rho$ reaches $\rho_0$, and the pressure vanishes.

\begin{figure}[!hbtp] 
\begin{center}
\includegraphics[width=10cm]{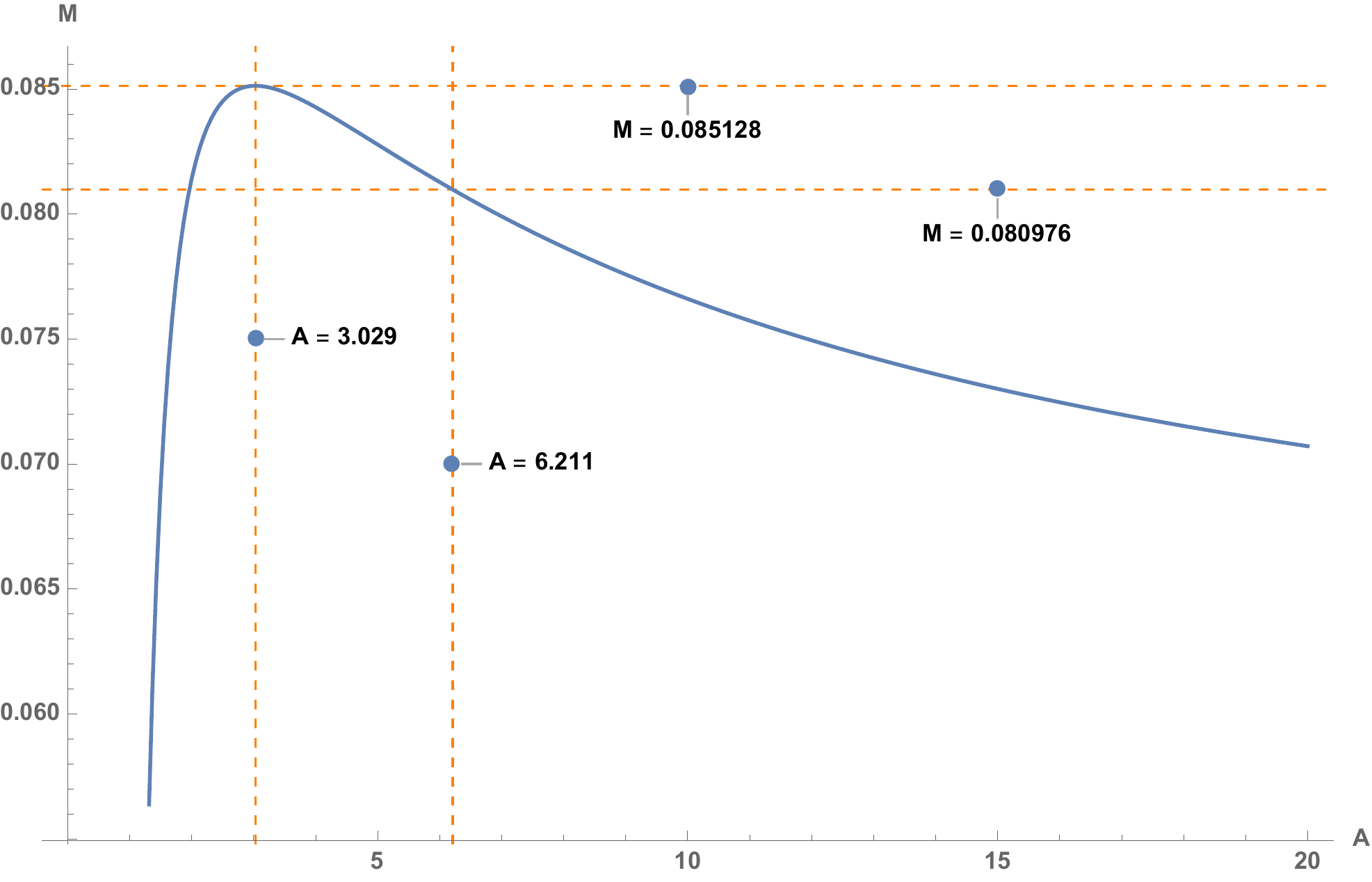}
\end{center}
\caption{Mass vs. central density parameter $A$ for solutions with the maximally stiff equation of state \eqref{mceos}.}
\label{fig1}
\end{figure}

\begin{figure}[!hbtp] 
\begin{center}
\includegraphics[width=10cm]{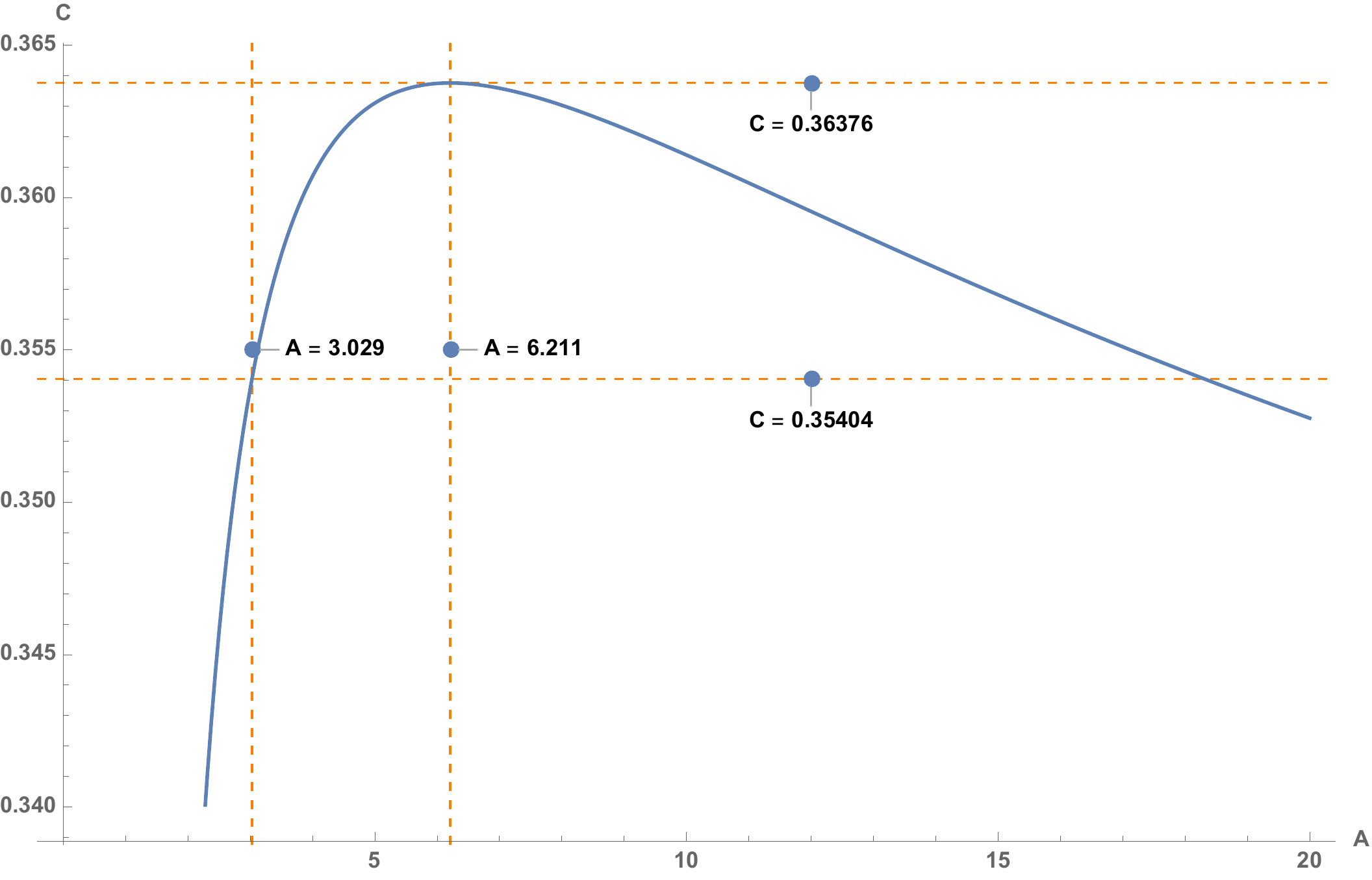}
\end{center}
\caption{Compactness vs. central density parameter $A$ for solutions with the maximally stiff equation of state \eqref{mceos}.}
\label{fig2}
\end{figure}

\begin{figure}[!hbtp] 
\begin{center}
\includegraphics[width=10cm]{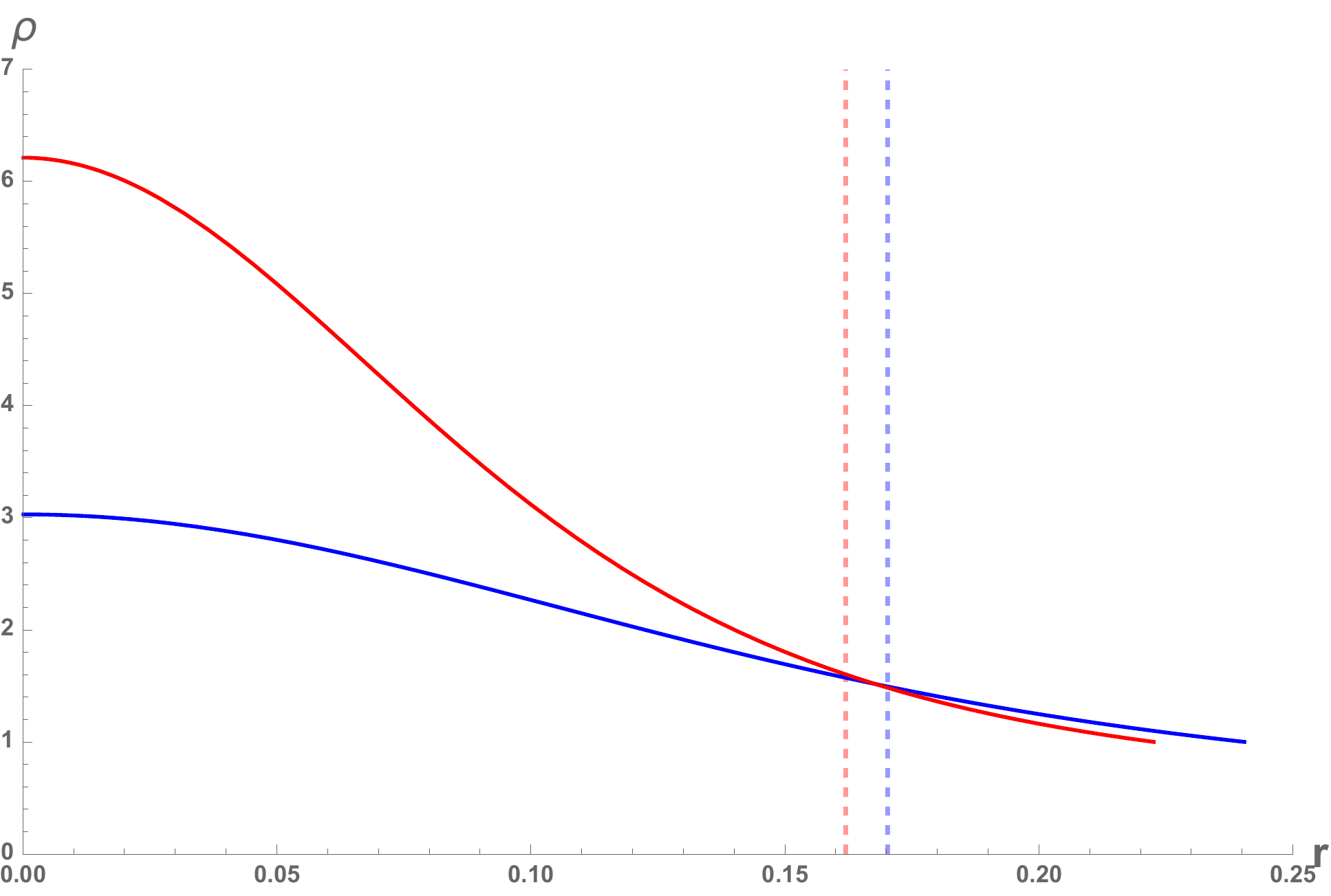}
\end{center}
\caption{Density profile for solutions with the maximally stiff EOS \eqref{mceos}.  Upper (red) curve corresponds to solution with maximal $C$ ($A=6.211$); lower (blue) curve corresponds to solution with maximal mass ($A=3.029$).  Vertical dotted lines mark the Schwarzschild radii corresponding to the total mass of the solutions.}
\label{fig3}
\end{figure}

\begin{figure}[!hbtp] 
\begin{center}
\includegraphics[width=10cm]{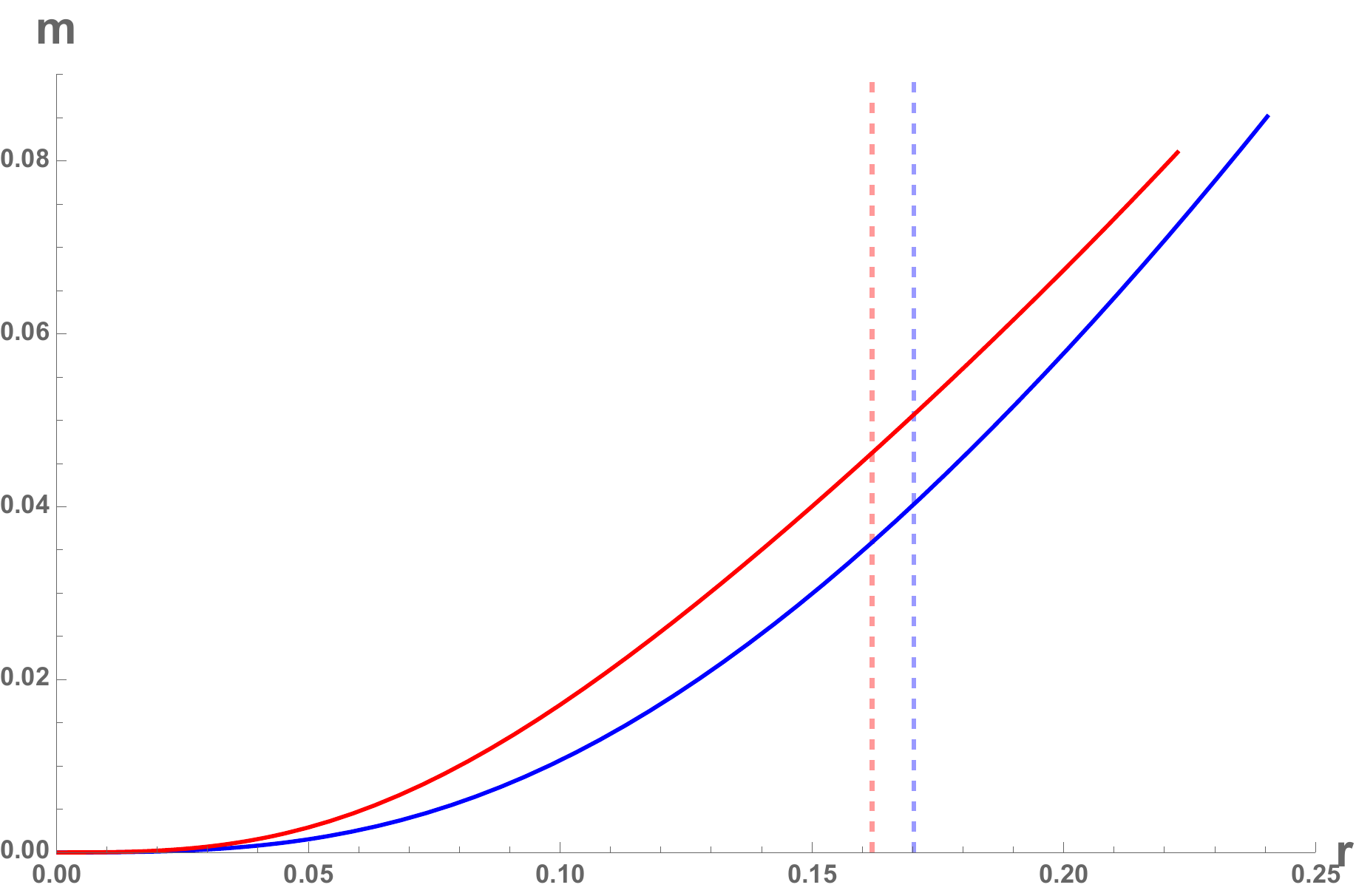}
\end{center}
\caption{Mass profile $m(r)$ for solutions with the maximally stiff EOS \eqref{mceos}.  Upper (red) curve corresponds to solution with maximal $C$ ($A=6.211$); lower (blue) curve corresponds to solution with maximal mass ($A=3.029$).  Vertical dotted lines mark the Schwarzschild radii corresponding to the total mass of the solutions.}
\label{fig4}
\end{figure}

It is worth describing this most compact case further, since these provide first candidate toy models for highly compact CQOs. The solutions may be found by specifying a central density $\rho_c=A \rho_0$ and integrating the TOV equations \eqref{TOVgen} outward from $r=0$.  
Plots of the total mass $M$ and compactness $C$ of the resulting solutions as a function of central density parameter $A$ are shown in figs.~\ref{fig1}, \ref{fig2}, and radial profiles of the density $\rho$ and mass within a given radius are shown in figs.~~\ref{fig3}, \ref{fig4}.   Linear stability properties change at the point where $dM/d\rho_c=0$, and so the higher density solutions are expected to be unstable.  This gives a maximum compactness $C_m=0.354$.  

The mass of the maximally compact solutions is determined in terms of the EOS parameter $\rho_0$ as
\beq\label{Mmax}
M_m= \frac{0.085\, c^4}{\sqrt {G^3 \rho_0}}\ ,
\eeq
and these have $A_m=3.029$.
So, if we want to describe such compact solutions with arbitrary masses, we need to consider a family of effective equations of state \eqref{mceos} with varying $\rho_0$.  The scaling transformation \eqref{Tscale} acts as $\rho_0\rightarrow \widetilde{\rho_0}= \lambda^2 \rho_0$, and correspondingly from \eqref{Mmax}
\beq
M_m\rightarrow \frac{M_m}{\lambda}\ ,
\eeq
as is seen with the Schwarzschild solution.

Since these solutions have compactness $C>1/3$, they are inside their light rings at $r=3M$.  This suggests that departures of the GW signal due to a merger of two of these from the signal  from two equal mass CBHs may be strongly obscured.  An important test of this would be to calculate the gravitational wave form arising from such a merger; a useful project would be to do so with numerical simulation.  However, there are also present practical limitations on such numerical simulations\footnote{We thank L. Lehner and D. Neilsen for discussions on this.}  and in particular the discontinuity in $dp/d\rho$ can be problematic for certain standard simulation routines.  For this reason, the next subsection will consider solutions with improved continuity properties; it will also turn out that certain properties depend on how the density drops to zero.

Another possible issue for solutions that lie inside their light rings is that of possible non-linear instability\cite{Keir:2014oka,Cardoso:2014sna,Cunha:2017qtt}, due to the trapping behavior of the effective gravitational potential of such a solution.  This remains a subject for further exploration, and of course relies on a classical analysis which may not apply to CQOs.
A pragmatic approach is to study evolution of binaries of such solutions, for example initiated not far from the orbital radius of plunge/merger; if instabilities are relevant on such timescales, they should be evident in the evolution.  In addition, such objects can become linearly unstable once they acquire  
spin\cite{Friedman:1978wla,Cardoso:2007az}.  If instabilities interfere with use of such solutions for testing GW departures, another approach is to adjust $\rho_c$ just to the point where a light ring or instability ceases to exist; the resulting solutions are still expected to be highly compact, and to still provide information about gravitational obscuration of signals from mergers of CQOs.

\subsection{Matched polytropes}

Improved continuity properties can be achieved by considering an EOS corresponding to the stiffest EOS \eqref{mceos} at high energy densities, but which then transitions to a polytropic EOS at an energy density $B\rho_0$, with $B>1$;
 \beq\label{mEOS}
p=
\begin{cases}
\rho-\rho_0 , & \text{if}\ \rho>B\rho_0 \\
      K(\rho/\rho_0)^\gamma, & \text{if}\ \rho<B\rho_0\ .
    \end{cases} 
\eeq
Here $K$ and $\gamma$ are fixed by the requirement that $p$ and its first derivative be continuous at $B\rho_0$, so that the EOS is $C^1$:
\beq
K= \frac{B-1}{B^\gamma}\rho_0\ ,
\eeq
\beq
\gamma=\frac{B}{B-1}\ .
\eeq

The improved continuity properties of \eqref{mEOS} suggest that the corresponding solutions are potentially useful, particularly for simulation with numerical GR.    These have similar features to those of the maximally compact case \eqref{mceos}, and in particular can achieve very high compactness.  They are also in some ways similar to models of neutron stars based on hybrid EOSs, but we will consider EOS parameters attaining much higher compactness than that of neutron stars, or of other simple EOSs such as a purely polytropic EOS.

It is important to emphasize that our $\rho$ is the energy density of the solution (that is, $T^0_0$ in Schwarzschild coordinates). In the modeling of neutron stars, the term `polytrope' is often used for simple toy EOSs where the pressure is a monomial of a `rest mass density' $\rho_m$ instead. In that context, the neutron star is composed of matter with a conserved quantum number, namely baryon number.  Then  parametrizing the rest mass density as $\rho_m = m_b n$, with $m_b$ a fixed average baryon mass and $n$ a conserved baryonic number density, allows for this additional conservation law to be accounted for when dynamics are turned on. The energy density is then related to this number density and pressure using an argument based on the first law of thermodynamics. For example, see \cite{Read:2008iy} on the use of piecewise-polytropic EOSs to model general NS EOSs, and \cite{NeDu} has considered matching the extremal EOS \eqref{mceos} onto such an EOS. Of course our EOSs, as toy models of quantum-gravitational corrections to CBHs, need not have any such conserved quantities (and surely should \textit{not} have exactly-conserved baryon number). Matching our extremal core EOS to an envelope EOS where the pressure is a monomial of the energy density  should allow for a unified numerical treatment of the two different regions of the solution.

\begin{figure}[!hbtp] 
\begin{center}
\includegraphics[width=10cm]{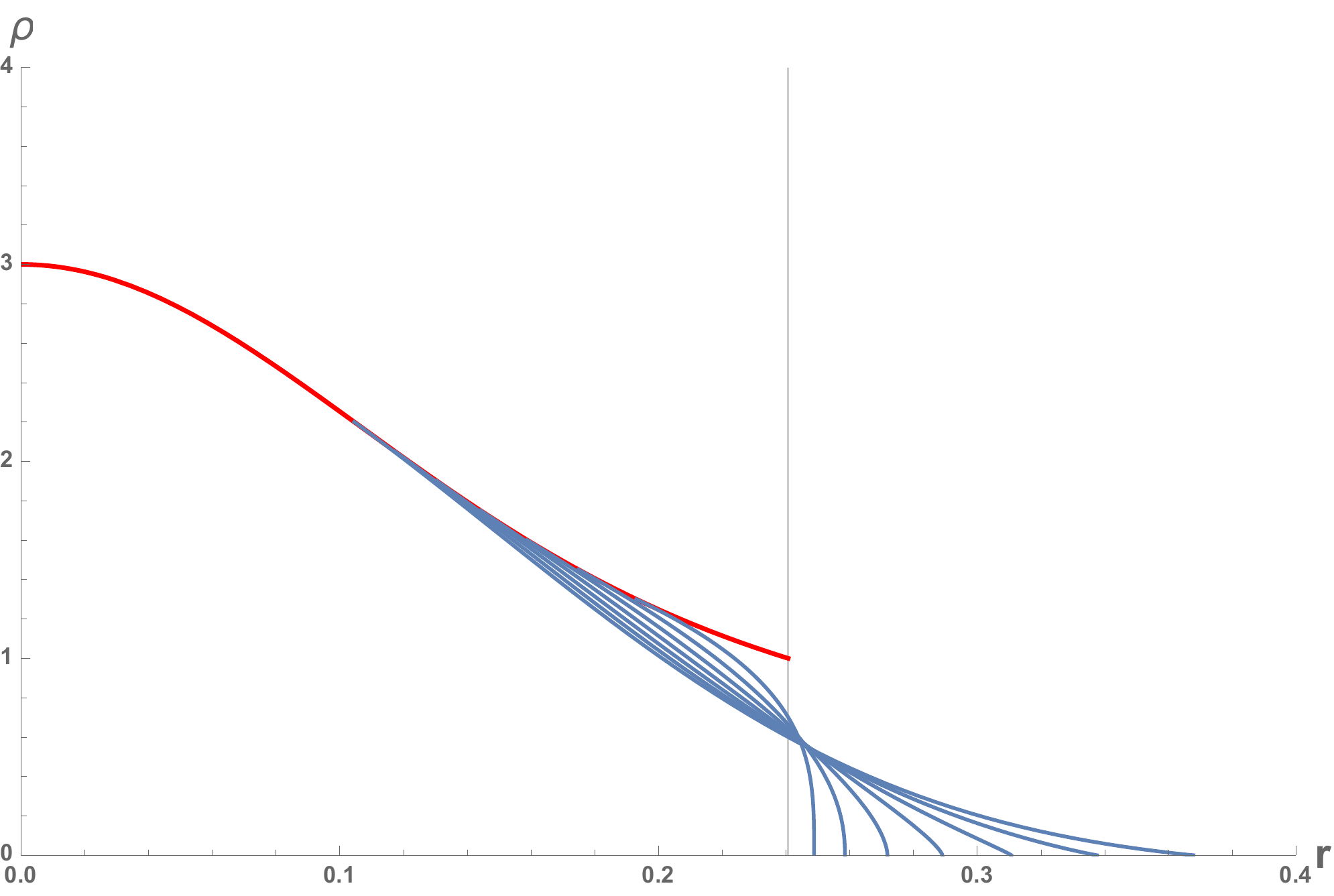}
\end{center}
\caption{Density vs. radius curves for matched solutions with EOS \eqref{mEOS}, with range of $B$ from 1.3 to 2.2 (lower, blue curves), and for corresponding extremal solution with EOS \eqref{mceos} (upper, red curve).}
\label{fig6}
\end{figure}

\begin{figure}[!hbtp] 
\begin{center}
\includegraphics[width=10cm]{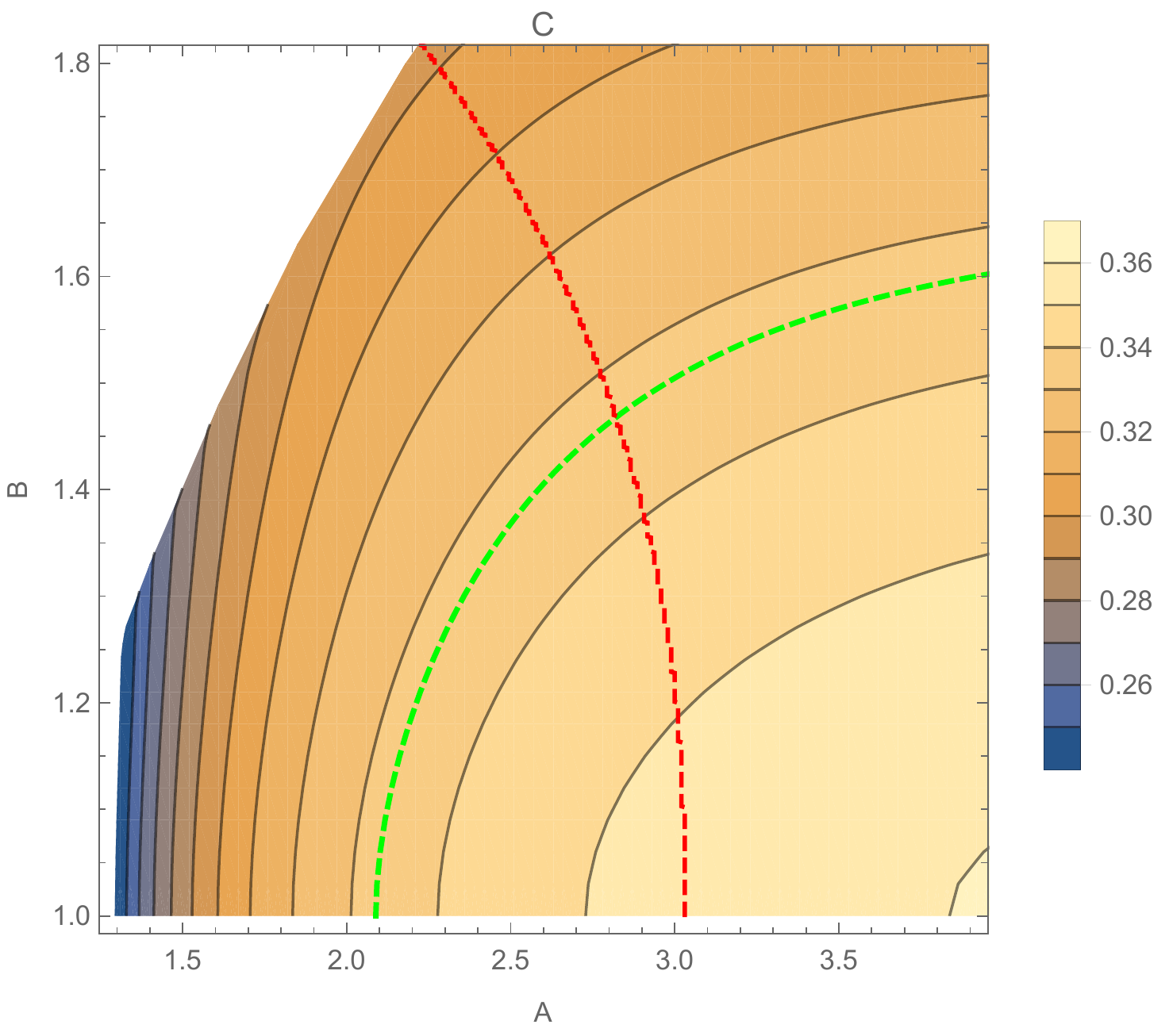}
\end{center}
\caption{Compactness of matched solutions with EOS \eqref{mEOS}, as a function of the parameters $A$ and $B$.  The red line denotes the stability boundary, where $dM/d\rho_c$ vanishes; solutions to the left  of this line are expected to be linearly stable.  The green line  demarcates compactness $C=1/3$.}
\label{fig5}
\end{figure}

Specifically, one can once again solve Einstein's equations together with the TOV equation.  The EOS is now determined by $\rho_0$ and $B$.  For a given EOS, spherically-symmetric, static solutions are determined by the central density $\rho_c=A \rho_0$.  These solutions have cores with the linear EOS in \eqref{mEOS}, and envelopes with the polytropic EOS in \eqref{mEOS}.  Example profiles of these solutions are shown in fig.~\ref{fig6}.  The compactness of these solutions, as a function of $A$ and $B$, is shown in fig.~\ref{fig5}.\footnote{While the matched polytropic EOS is well-defined up to the line $A=B$, as $B \rightarrow A$ the solutions become predominantly composed of the polytropic envelopes, which leads to lower compactness and also to longer numerical integration time. In Figure \ref{fig5} the upper cutoff corresponds to solutions where the outer radius of the solution was at three times the radius of the inner boundary, so the omitted solutions are $> 3/4$ envelope in terms of radial extent. As is evident, the plotted region contains the parameter space of greatest interest, namely where the solutions are both ultracompact and linearly stable.}  Clearly solutions exist, for a range of $B$, with compactness $C>1/3$.  The maximal compactness for such linearly-stable solutions is obtained in the (singular) limit of the EOS, $B\rightarrow1$.  As before, we expect the mass of such solutions to scale as $M\propto 1/\sqrt{\rho_0}$.

The $C^1$ EOS \eqref{mEOS} thus furnishes solutions that may possibly be more suitable for simulation of inspiral and merger via numerical GR methods.  We next turn to discussion of some general aspects of such evolution.

\section{Gravitational wave tests of compact quantum structure}

If consistency with quantum mechanics dictates that CBHs are replaced by compact quantum objects with  different properties, it is important to understand the possible sensitivity of gravitational wave observations, with LIGO/VIRGO or LISA, to this new structure.  Since we don't yet have a first-principles description of such objects that follows from a complete quantum theory of gravity, a first step is to find approximate or effective characterizations of the deviations from CBH behavior, and try to determine how these deviations might manifest themselves in modifications to GW signals.\footnote{In the regime where departures from GR are small, including inspiral and plunge up to the point of significant interactions, useful approaches may include the post-Einsteinian framework\cite{Loutrel:2014vja} or effective field theory\cite{Porto:2016pyg}.}    Indeed, precisely because we don't yet have a complete quantum theory describing black hole-like objects, any observational (or experimental) evidence for deviations from GR could provide extremely useful guidance, and should be searched for by all available means.  A way to begin to understand and characterize such possible deviations is by introducing parameterizations of structure and dynamics of objects replacing CBHs, and investigating the GW sensitivity and how it depends on those parameterizations.  This can help focus the  search for possible departures from GR.  In cases where we can provide tractable models for the new dynamics, these can be investigated with numerical GR methods.

We can distinguish three levels of detail in characterizing behavior of candidate CQOs:  their description in terms of simple effective parameters, modeling of such objects in toy models or effective dynamical descriptions, and a complete description in a more fundamental theory of quantum gravity.  

\subsection{Effective parameters}

The first level of detail focuses on the sizes of certain key parameters\cite{SGobswind,SGGW,SG-nature}.  For example, the departure of CQO structure from that of a CBH may extend to a characteristic radius $R_a=R+\Delta R_a$ outside the horizon radius $R$ of the CBH with the same mass.\footnote{For simplicity we ignore spin dependence, but in the more general spinning case the ``height" $\Delta R_a$ of the gravitational atmosphere may vary {\it e.g.} by an order one amount as a function of angle.} The CQO structure may also be ``hard" or ``soft."  A characteristic of this distinction is the typical spatial variation scale, or momentum transfer scale for particles as they scatter from the structure; a simple parameterization of this is as $1/L$, where $L$ is a length scale describing the variation of the structure.  Obviously $L\lesssim \Delta R_a$, and $L$ may be much smaller, {\it e.g.} a microscopic size.  Another characteristic is the strength $\cal A$ of the departures from GR; a benchmark for this is the amplitude for scattering of excitations from a CQO to depart from scattering from a CBH.  A fourth parameter is the timescale $T_q$ at which a CQO exhibits behavior departing from that of a CBH.  In order for CQO dynamics to resolve quantum problems with BHs, this timescale is expected to lie in the range  
from $\sim R\log R$ -- a short timescale in astrophysical terms -- to $\sim R^3$, much longer than the age of the Universe for stellar-sized or larger BHs.  Other potentially important parameters include tidal deformability, quantified by the Love numbers, and absorption cross sections, which can depend on wavelength.

These parameters play an important role in governing expected departure from GR predictions for GW signals.  For example, departures will only be found for CQOs with age longer than $T_q$.  For such CQOs, changes in Love numbers can lead to small departures in the GW signal from inspiral.  Then, as CQOs plunge to merger, important departures from GR behavior can be expected at CQO separations $\sim 2(R+\Delta R_a)$, where the structures come into contact.  This expectation holds for strong departures, $\cala \sim 1$, but not necessarily if $\cala \ll 1$, and the departures are expected to depend on other efective parameters such as $L$.  A model for such departures\cite{SG-nature} is merger of neutron stars, which give an example of large deviations in GW signals from those of CBHs, due to hard, $\calo(1)$ structure outside the would-be horizon.

For example, a typical massive remnant scenario has hard ($L$ microscopic), $\cala\sim 1$ structure, but without a more detailed model, $\Delta R_a$ could range from microscopic values to $\gtrsim R$.  Only in the latter case would one expect significant modifications to GW signals.\footnote{This assumes that massive remnants merge to form BHs, and that $T_q\gtrsim R\log R$; exceptions will be discussed below.}  Love numbers and absorption cross sections also depend on the details of the model.

The soft gravitational atmosphere case has $\Delta R_a\sim L\sim R$, or more generally $\Delta R_a\sim R^q$, $L\sim R^p$, for some $p,q>0$, to achieve ``soft" scales for large CQOs.  There are two variants, the strong one\cite{NVNLT} with $\cala\sim 1$, and the weak one\cite{NVU}  with $\cala\sim 1/\sqrt N$, where $N$ characterizes the large number of internal states.  The strong case is expected to modify the GW signal, but also the weak case can lead to modified absorption cross sections\cite{SGcross} of GWs with wavelength $\sim R$,  and thus also yield departures in gravitational wave signals.

The EOS-based models considered in the preceding section also illustrate possible such parameters.  The departures from the Schwarzschild metric extend over scales $\Delta R_a\sim R$, as can be seen from figs.~\ref{fig3}, \ref{fig4}.  Likewise, the characteristic variation scale of the geometry of these solutions is  $L\sim R$; the gravitational departures are ``soft" in this sense.  Note, however, that if the EOS \eqref{mceos} or \eqref{mEOS} gives an effective description of fluctuations of the geometry or other fields at microscopic distances, then in this more fundamental dynamics $L\ll R$.  This could then reveal itself in the interaction properties of two such objects.  This can be illustrated by the analogous collision of two neutron stars.  The macroscopic geometries of the neutron stars vary on scales $L\gtrsim R$.  However, when the surfaces of neutron stars approach in a collision, the interactions of the neutron condensates of the two stars are important, and are characterized by hard scales $L\ll R$.  Absorption into or scattering from such objects, if it involves interactions other than with the macroscopic gravitational field, may also be characterized by scales $L\ll R$.  With such hard structure, the models of section \ref{Models} behave like massive remnants, although if one considers only their average gravitational field they behave more like soft gravitational atmospheres.

\subsection{Love numbers}

\begin{figure}[!hbtp] 
\begin{center}
\includegraphics[width=10cm]{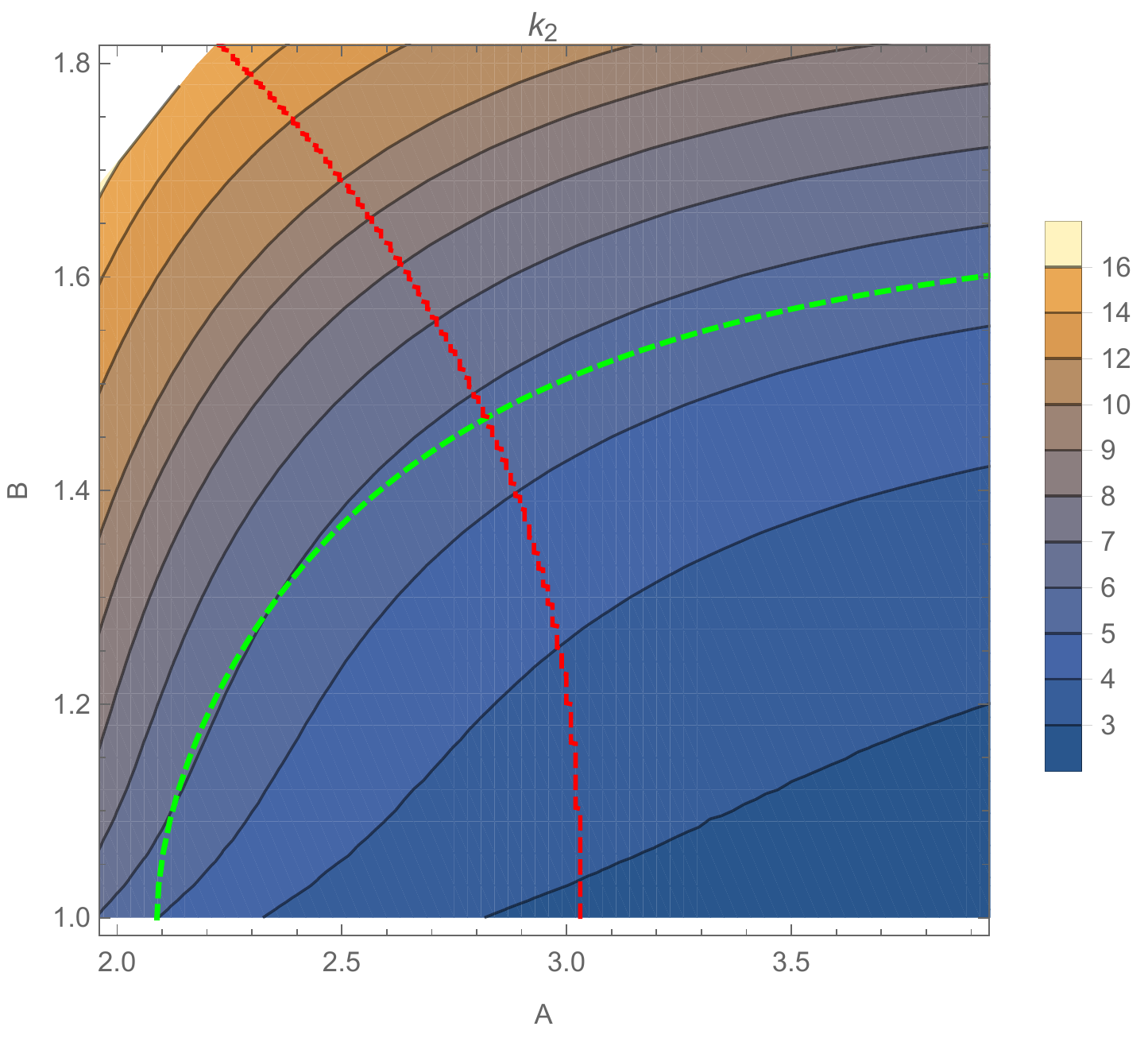}
\end{center}
\caption{Love number $k_2$ of matched solutions with EOS \eqref{mEOS}, as a function of the parameters $A$ and $B$.}
\label{fig7}
\end{figure}

Love numbers, characterizing tidal deformability, are particularly relevant parameters for investigating deviations during inspiral.  
With an EOS such as \eqref{mceos} or \eqref{mEOS}, one may calculate these.  For the case of the matched EOS \eqref{mEOS}, a plot of the numerically-calculated Love number $k_2$ is shown in fig.~\ref{fig7}.  We calculate the $\ell = 2$ polar tidal Love number (TLN) using the method outlined in \cite{Hinderer:2007mb},\footnote{For a general analysis of $\ell \ge 2$, see \cite{1967ApJ...149..591T}.}; it is defined as $k_2=3\lambda M^{-5}/2$, where $\lambda$ is the proportionality constant between the quadrupole moment $Q$ and tidal field $\cal E$, $Q_{ij}=-\lambda {\cal E}_{ij}$.  The definition of $k_2$ is
normalized to the mass of the objects (rather than the radius), as proposed in \cite{Cardoso:2017cfl}, since generic models of CQOs need not have well-defined radii.

We utilize \texttt{Mathematica} \cite{Mathematica} to numerically solve the TOV equations for the matched polytropes and then to calculate the first-order $\ell=2$ response to an external quadrupolar tidal field. This allows us to calculate the TLN $k_2$ which encodes information about the internal structure of the object. As explained in \cite{Hinderer:2007mb}, this constant is sensitive to the boundary conditions imposed at the outer radius of the star $R$. As seen in Figure \ref{fig7}, values of $k_2$ range from $\sim 2-6$ in the region of parameter space in which the solutions are ultracompact and linearly stable.
We will comment below in \ref{inspiral} on the physical effects of $k_2$ during merger, and its measurability.  (Discussion of uncertainty in these calculations of $k_2$ appears in the appendix.)

\begin{figure}[!hbtp] 
\begin{center}
\includegraphics[width=10cm]{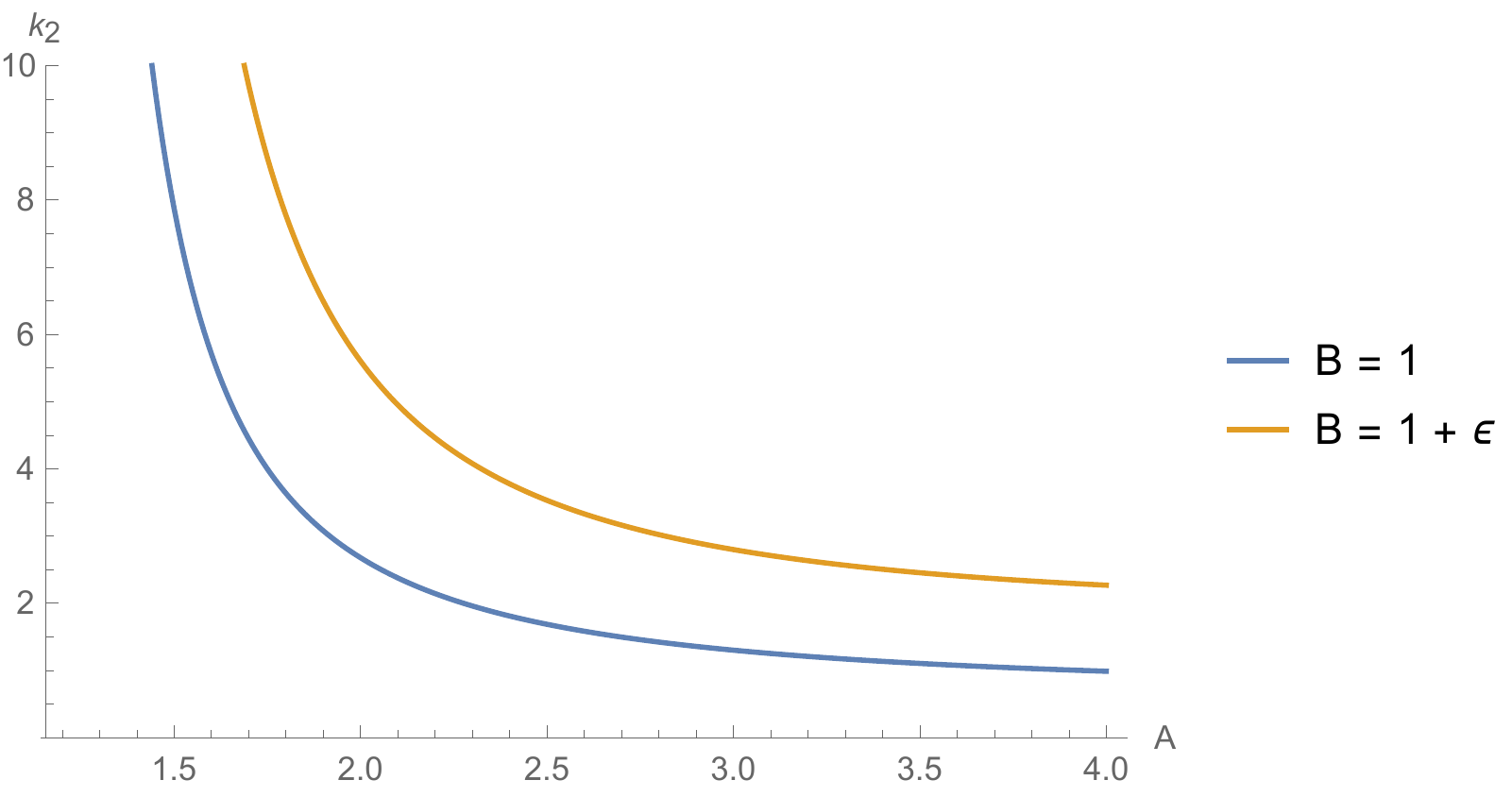}
\end{center}
\caption{Tidal Love numbers of the extremal solutions with EOS \eqref{mceos} (``$B=1$"), compared to those with the matched EOS \eqref{mEOS}, in the limit approaching $B=1$. }
\label{lovediff}
\end{figure}

It is important to note that while our matched polytropes have a continuous family of EOSs parametrized by $B$, the solutions of the TOV equations with these EOSs are not continuous in the limit $B \rightarrow 1$. This is visible already in Figure \ref{fig6}, where one can see that the outer radius of the extremal EOS $B=1$ has $\rho(R) = \rho_0$, whereas the energy density falls to zero at the outer radius for any $B\neq1$. 
As the TLN is sensitive to the boundary conditions at $r = R$, this difference in boundary behavior significantly affects the TLN calculation.
In Figure \ref{lovediff} we show the TLNs of the extremal solutions and of the matched polytropic solutions in the limit $B \rightarrow 1$ and find that the presence of the envelope roughly doubles the TLN, even in the limit of vanishing size.  This serves as one illustration of  
the necessity of including parameters beyond $\Delta R_a$ in an effective description of CQOs, as despite having the same compactnesses, the macroscopic, observable effects of these objects in inspiral differ significantly.

\subsection{Evolution: effective models and expectations}

The effective parameters described above may be useful in characterizing the signal departures from that of CBHs.  For a more precise determination of these departures, one needs to consider the second level of detail, using models for dynamics of CQOs.  Details of dynamics can be important; for example, 
even in the massive remnant scenario with $\Delta R_a\sim R$, GW signal departures may be suppressed through gravitational obscuration (absorption into the final BH).  One approach to such dynamics is via evolving solutions with effective EOSs.

Thus, for example, we would like to understand departures from CBH behavior in the gravitational wave signal arising from the collision of two solutions with effective EOS  \eqref{mceos} or \eqref{mEOS}.  Ideally, this could be addressed via numerical simulation of the inspiral and merger of two such solutions.  In turn, understanding the signal departure then can give further insight into the roles of the effective parameters $\Delta R_a$, $L$, and $k_2$, in governing gravitational wave signals from other models for CQOs.

In fact, given the parameters we have discussed, we can describe some anticipated features of this evolution and the signal departures.  

\subsubsection{Inspiral}\label{inspiral}

To begin with, consider the inspiral phase.  Here, the principal modification to the GW signal is expected to arise from the tidal deformability of the compact objects, as parameterized by the Love number $k_2$, and there is a clear and precise connection between this parameter and the predicted signal deviation. 

This TLN affects the signal at fifth post-Newtonian order and adds linearly to the phase of the waveform in frequency-space as
\begin{equation}
\tilde{h}(f) = \mathcal{A}(f) e^{i \left[\psi_{PP}(f) + \psi_{TD}(f)\right]},
\end{equation}
where $f$ is the frequency, $\mathcal{A}$ is the amplitude, $\psi_{PP}$ is the would-be phase for a merger of point particles, and $\psi_{TD}$ the effect from the tidal deformability of the merging objects. This is related to the TLN $k_2$ as \cite{Flanagan:2007ix}
\begin{equation}
 \psi_{TD}(f) = - \frac{3}{8} \frac{v^5}{\mu {\cal M}^4} \left[ \left( 11 \frac{M_2}{M_1} + \frac{\cal M}{M_1} \right) M_1^5 k_2^{(1)} + 1 \leftrightarrow 2 \right],
\end{equation}
where $v = (\pi {\cal M} f)^{1/3}$ is the inspiral velocity, $\cal M$ and $\mu$ are the total and reduced masses of the binary, and $M_i$ and $k_2^{(i)}$ are the individual masses and TLNs. This effect has been used to constrain the tidal deformability of neutron stars \cite{TheLIGOScientific:2017qsa}, which places nontrivial constraints on the neutron star EOS already from the first observation of a binary merger (see {\it e.g.} \cite{Annala:2017llu}).

The TLNs of static CBHs vanish exactly. This interesting result was first noted in \cite{Fang:2005qq}, and further studied in \cite{Binnington:2009bb, Damour:2009vw, Gurlebeck:2015xpa, Poisson:2014gka, Pani:2015hfa, Landry:2015zfa}. Contrastingly, other compact objects generically have non-zero tidal deformability, which makes this effect a good discriminating feature (as suggested in \cite{Porto:2016zng}).
Ref.~\cite{Cardoso:2017cfl} used Fisher matrix methods to systematically compute the sensitivity of GW detectors to this parameter for compact objects, and finds broadly that Advanced LIGO at design sensitivity may be able to constrain TLNs down to $k_2 \sim 100$ across a broad range of binary masses; the analysis of \cite{Sennett:2017etc} used similar methods and is in broad agreement.\footnote{The effect of tidal deformations in modifying the ``contact frequency" has also been used to constrain some simple models of boson stars (which are not ultracompact) as being the sources of LIGO's `black hole' mergers \cite{Johnson-McDaniel:2018uvs}, though this does not take advantage of all of the constraining power of the effects of tidal deformability.} We thus use this number as a benchmark for near-term sensitivity to tidal effects.  Neutron star models have tidal Love numbers of this magnitude.
However, it is clear from Figure \ref{fig7} that the tidal effects of the present very compact CQO models are far too small to be measurable with Advanced LIGO.  In \cite{Cardoso:2017cfl}, a study of compact objects with $\Delta R_a \approx l_P$ found $k_2 \approx \mathcal{O}(10^{-3})$, together with a universal logarithmic dependence of the TLNs on the location of the surface in the $\Delta R_a \to 0$ limit, in agreement with $k_2 = 0$ for CBHs.\footnote{For work to distinguish tidal parameters for neutron stars see \cite{Annala:2017llu,Landry:2018prl}.} In order to distinguish a CQO merger from a CBH merger by its tidal deformability alone, space-based detectors appear necessary. The possibility of observing strong gravity effects with these machines should serve as additional motivation for their construction, as well as for further detailed study of their capabilities\cite{Maselli:2017cmm, Cardoso:2017cfl, Maselli:2018fay}. Ref.~\cite{Cardoso:2017cfl} shows that current designs of future space-based GW detectors may be able to do up to two orders of magnitude better in optimistic detection scenarios, though it has recently been pointed out in \cite{Maselli:2017cmm} that the effect of tidal heating dominates for LISA binaries. (Ref.~\cite{Maselli:2018fay} suggests even greater LISA sensitivity to highly-spinning objects.)  It may also be possible to extract signatures of departures from CBHs by ``stacking'' signals from multiple events, as in \cite{Yang:2017xlf}.\footnote{It also may be worth using such methods with binary neutron star templates even at high masses, to explore for possible departures.  We thank L. Lehner for pointing out the possible role of such stacking.}

The structure of objects of moderate compactness can also significantly shift the end of inspiral -- objects can for example begin to interact before they reach what would be the ISCO for CBHs\cite{Lai:1996sv,Giudice:2016zpa}.  However, while further investigation is warranted, such effects are not expected to be important for objects with compactness in the range $C\gtrsim 1/3$.  Such objects are expected to have important modifications due to later interactions after they have entered the plunge phase, as they begin to merge.  This observation and the preceding challenges emphasize the importance of moving past inspiral and gaining detailed understanding of the behavior of CQOs during plunge, where their deviations from CBHs will be more pronounced.

\subsubsection{Plunge and Merger}

Larger deviations are expected when the model CQOs approach the point where they merge.  For equal mass CQOs with mass $M$, inspiral transitions to plunge at a separation 
$d\sim 24M$, corresponding to the mutual innermost stable circular orbit.  So objects with compactness $\gtrsim1/12$ will have a plunge phase, terminated by this merger at separation $d\sim 2R_a$.  The GW signal departs from that of a CBH collision both because the objects encounter the gravitational field modifications, and because they increasingly disrupt each other's structure.   Neutron star collisions, for example as modeled in \cite{Palenzuela:2015dqa}, furnish an example of some of these possible modifications\cite{SGGW}.  In neutron star simulations, one finds a significant increase in the GW amplitude when the stars begin to merge (see {\it e.g.} fig.~2 of \cite{Palenzuela:2015dqa}), but a decrease in the emitted power spectrum;  a scaled up version of the examples of \cite{Palenzuela:2015dqa}, using the scaling transformation \eqref{scalexm}, \eqref{pertrescale}, for example, already appears inconsistent with LIGO data.  

While similar features are expected for simple CQO models, such as those given above, CQOs may be much more compact than neutron stars, which have compactnesses $C\lesssim 1/5$.  In the case where the compactness approaches, or exceeds, $C=1/3$, one expects some signal  obscuration, as noted above, since the departures due to merger are generated ``deep in the gravitational potential."  This is not necessarily a sharp boundary.  Indeed, such a configuration has a significant angular momentum; this means for example that the prograde and retrograde light rings of the ultimate object have different radii.  If one assumes that most of the perturbations are prograde, the fact that the prograde light orbits  tend to cluster towards $r\sim M$ for high spin then suggests the possibility of reduced obscuration. One also expects the GW signal to depend on other details of the model for the colliding objects, including how their structure interacts.  

Given the uncertainties, a particularly interesting project would be  to simulate the CQO models described above via numerical methods.  It appears that already some important questions -- such as the role of obscuration -- can begin to be addressed even in the simple models with effective EOS given in section \ref{Models}.  An example of a concrete question is whether recent GW detections -- {\it e.g.} the first LIGO detection of a merger of two $\sim 30 M_{\odot}$ objects -- are capable of distinguishing a CBH merger from a merger of objects governed by the EOS \eqref{mEOS}, if the EOS parameters are such that the solutions are highly compact, $C\gtrsim 1/3$. We hope to see such models studied via numerical GR in the near future.  As an intermediate step, it may also be possible to formulate hybrid waveforms which approximately capture important aspects of the GW signatures for model CQOs, by matching tidally-corrected inspiral waveforms  to parameterizations of waveforms for plunge and merger, and then to the quasinormal regime\cite{LehnPC}.

\subsection{Quantum dynamics and expectations}

We would of course like to be describing CQO evolution within the context of a complete quantum theory of quantum gravity, providing a consistent description of quantum analogs of black holes, but the field has not sufficiently advanced.  The models described above may supply some insight into how sensitive observations are to this more complete quantum dynamics.  However, a key question is how accurately the models capture important features of the full quantum evolution.

As was noted in section \ref{motiv}, there are multiple contenders for scenarios for the quantum completion of black hole evolution.  

In the massive remnant category, one possibility that has been repeatedly considered is that of a massive remnant with $\Delta R_a \ll R$.  Firewalls\cite{AMPS} fit in this category.  If there is a viable fuzzball scenario, it is not known if it produces $\Delta R_a\sim R$ or $\Delta R_a \ll R$, but the latter has been suggested\cite{Guo:2017jmi} and would fit in this category.  Such objects have also been considered, and given the name ``ClePhOs," in \cite{Cardoso:2017njb}.  However, there is so far no dynamical theory or effective model that produces any of these objects, and allows study of their evolution, and an important question is whether such configurations can exist in a consistent description.  Possibly, such highly-compact solutions could be modeled with an anisotropic stress tensor\cite{Raposo:2018rjn}, avoiding the Buchdahl bound \eqref{buchbd}, with a stress tensor violating other conditions ({\it e.g.} gravastars), or by other means.  If such a highly compact configuration does give the correct physics, it is important to develop a consistent dynamical description for it, even if it is in an effective model.  

If such configurations {\it were} physical, the next question is whether observations would be sensitive to their features.  The case $\Delta R_a\ll R$ appears difficult, although there is possible sensitivity through TLNs\cite{Maselli:2018fay}.  Electromagnetic images are determined by trajectories of photons in the vicinity of the light ring, and so are not necessarily sensitive to such a configuration.  Likewise, given the preceding discussion surrounding obscuration, it appears even less likely that the collision of two such highly-compact objects would substantially alter gravitational wave signals.

Of course, one proviso in this is if such objects coalesce to form another such object that does not behave like a black hole.  There is no compelling reason for this to happen; if such objects exist, they are plausibly expected to  coalesce to form black holes that {\it later} transition to a ClePhO, after a time at least $\sim R\log R$.  But, if a ClePhO {\it were} to form promptly, at times $\ll R\log R$, then it could be possible for the surface of the object to reflect gravitational waves, and produce echoes\cite{Cardoso:2016rao,Cardoso:2016oxy,Abedi:2016hgu} of its formation.  However, such a scenario obviously requires a sequence of non-trivial assumptions.

A promising alternative to such a hard scenario is that of a soft gravitational atmosphere\cite{NVNLT,NVU}.  This can be described in terms of metric perturbations $h$ (or, more generally, other field perturbations) that depend on the quantum state of the black hole.  It is not clear how a soft atmosphere scenario could have dynamics summarized in terms of an effective EOS.  The strong version of \cite{NVNLT} has perturbations with size $\langle h\rangle \sim 1$, and thus could possibly yield an observational signal, but the more complete nonlinear dynamics is needed for its prediction.\footnote{Possible electromagnetic signals are discussed in \cite{GiPs}.}  

The  weak scenario of \cite{NVU} involves perturbations such that $\langle h\rangle$ can be exponentially small in the black hole entropy, representing a smaller departure from a GR-based description.    While this may seem more plausible, it na\"\i vely
looks problematic for observation.  However, due to the large number of black hole internal states, quantum scattering and absorption cross sections from such an object can receive $\calo(1)$ corrections, which are particularly important for modes with wavelength $\sim R$; this follows from an extension of the estimates for transition rates given in \cite{NVU} to the case with scattered radiation\cite{SGcross}.
 If these provide the leading quantum corrections, that can produce an observational effect on gravitational wave modes, which in a merger typically have wavelengths characterized by the same scale.  It may be possible to parameterize such absorption effects and analyze their effects on the gravitational signal.  This may be even more straightforward in the case of extremal mass ratio binaries where absorption into the larger mass object can be parameterized and its effect on the coalescence and radiation inferred\cite{OSullivan:2014ywd,OSullivan:2015lni};\footnote{We thank S. Hughes for a discussion on this question.} these investigations are left for future work.

\section{Acknowledgements}
We thank R. Adhikari, N. Craig, J. Hartle, D. Holz, G. Horowitz, S. Hughes,  L. Lindblom,   L. Stein, and particularly D. Neilsen, for valuable conversations.  
We also thank L. Lehner for particularly helpful conversations, and for comments on a preliminary version of this paper.
SG thanks the CERN theory group, where part of this work was carried out, for its hospitality.
The work of SG was supported in part by the U.S. Department of Energy, Office of Science, under Award Number {DE-SC}0011702, that of SK was supported in part by the U.S. Department of Energy, Office of Science, under Award Number {DE-SC}0014129, and the work of GT was supported in part by  NSF Grant PHY-1801805 and by a UC MEXUS-CONACYT Doctoral Fellowship.

\appendix
\section{Uncertainty in $k_2$ calculations}

The main source of uncertainty in our calculation of $k_2$ comes from the determination of the outer radius $R$, which is defined as the location at which the pressure vanishes. Due to the stiffness of our envelope EOS, the energy density falls quite steeply near the outer radius - especially as $B \rightarrow 1$, as seen in Figure \ref{fig6}. As a result, our numerical implementation of the TOV equations turns into a `stiff system of differential equations', which causes \texttt{Mathematica}'s numerical solution to fail at $R_{\text{end}}$, some small distance before the true outer radius $R-R_{\text{end}} \lll R$, when the pressure has not yet reached precisely zero.

\begin{figure}[!hbtp] 
\begin{center}
\includegraphics[width=10cm]{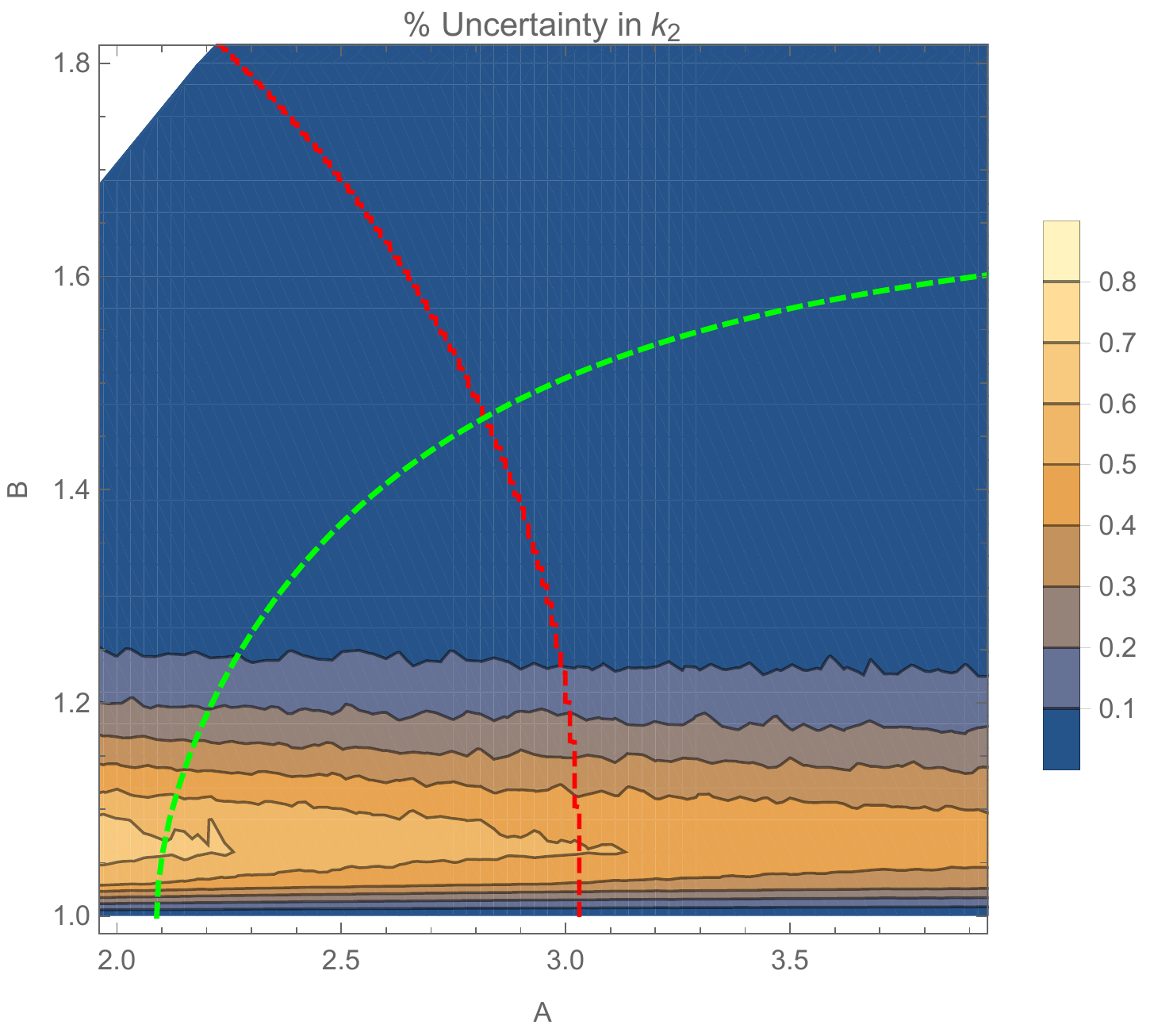}
\end{center}
\caption{}
\label{loveunc}
\end{figure}

 For a rough sense of the uncertainty this gives our calculation, we extrapolate the quickly-falling pressure past the point at which the solution fails and find where it hits zero. We assign an outer radius uncertainty $\Delta R$ to be the difference between this radius and where the solution stopped at $R_{\text{end}}$. We then assign an uncertainty to the TLN of $\Delta k_2 \equiv k_2'(R_{\text{end}}) \Delta R$, where $k_2'$ is the derivative of the TLN with respect to the outer radius at which it is calculated. While $\Delta R$ is very small, the TLN is sensitive exactly to the behavior of the solution at the boundary, which changes rapidly, so it is not obvious a priori that the resulting uncertainty should be negligible. Nevertheless, in Figure \ref{loveunc} we plot the percent uncertainty $\frac{k_2'(R_{\text{end}}) \Delta R}{k_2(R_{\text{end}})} \times 100$, which suggests that the calculation is under reasonable control everywhere.

\bibliographystyle{utphys}
\bibliography{stronggw}

\end{document}